\documentclass[12pt]{iopart}

\usepackage{graphicx}
\usepackage{hyperref}
\usepackage{xspace}
\usepackage{fontawesome5}
\usepackage{xcolor}

\newcommand{\runII}{\ensuremath{\mathrm{Run~2}}\xspace}

\newcommand{\gaudi}{\ensuremath{\mbox{\sc Gaudi}}\xspace}
\newcommand{\gauss}{\ensuremath{\mbox{\sc Gauss}}\xspace}
\newcommand{\boole}{\ensuremath{\mbox{\sc Boole}}\xspace}
\newcommand{\pythia}{\ensuremath{\mbox{\sc Pythia8}}\xspace}
\newcommand{\evtgen}{\ensuremath{\mbox{\sc EvtGen}}\xspace}
\newcommand{\geant}{\ensuremath{\mbox{\sc Geant4}}\xspace}
\newcommand{\lamarr}{\ensuremath{\mbox{\sc Lamarr}}\xspace}
\newcommand{\LbLcmunu}{\ensuremath{\Lambda_b^0 \to \Lambda_c^+ \mu^- \bar \nu_\mu}\xspace}
\newcommand{\LcpKpi}{\ensuremath{\Lambda_c^+ \to p K^- \pi^+}\xspace}

\newcommand{\orcid}[1]{\href{https://orcid.org/#1}{\textcolor[HTML]{A6CE39}{\faOrcid}}}

\begin{document}

\title{Lamarr: LHCb ultra-fast simulation based on machine learning models 
       deployed within Gauss}


\author{Matteo Barbetti$^{1,2}$ on behalf of the LHCb Simulation Project}

\address{
$^1$ Department of Information Engineering, University of Firenze,\\
\,\,\, via Santa Marta 3, Firenze (FI), Italy\\
$^2$ Istituto Nazionale di Fisica Nucleare, Sezione di Firenze,\\
\,\,\, via G. Sansonse 1, Sesto Fiorentino (FI), Italy
}






\ead{\href{mailto:matteo.barbetti@cern.ch}{Matteo.Barbetti@cern.ch}}

\begin{abstract}
About 90\% of the computing resources available to the LHCb experiment has been spent to produce simulated data samples for Run 2 of the Large Hadron Collider at CERN. The upgraded LHCb detector will be able to collect larger data samples, requiring many more simulated events to analyze the data to be collected in Run 3. Simulation is a key necessity of analysis to interpret signal, reject background and measure efficiencies. The needed simulation will far exceed the pledged resources, requiring an evolution in technologies and techniques to produce these simulated data samples. In this contribution, we discuss \lamarr, a \gaudi-based framework to speed-up the simulation production parameterizing both the detector response and the reconstruction algorithms of the LHCb experiment.
Deep Generative Models powered by several algorithms and strategies are employed to effectively parameterize the high-level response of the single components of the LHCb detector, encoding within neural networks the experimental errors and uncertainties introduced in the detection and reconstruction phases. Where possible, models are trained directly on real data, statistically subtracting any background components by applying appropriate reweighing procedures. 
Embedding \lamarr in the general LHCb \gauss Simulation framework allows to combine its execution with any of the available generators in a seamless way. The resulting software package enables a simulation process independent of the detailed simulation used to date.
\end{abstract}

\section{Introduction}
\label{sec:intro}
The LHCb detector~\cite{LHCb:2008vvz, LHCb:2014set}, originally designed to study particles containing $b$ and $c$ quarks produced at the Large Hadron Collider~(LHC), is a single-arm forward spectrometer covering the pseudorapidity range $2 < \eta < 5$.
The detector includes a high-precision tracking system providing measurements of the momentum $p$ of charged particles and the minimum distance of a track to a primary vertex~(PV), namely the impact parameter~(IP).
LHCb is also equipped with a highly performing particle identification~(PID) system capable of distinguishing photons, electrons, long-lived hadrons, and muons, combining the response of two ring-imaging Cherenkov~(RICH) detectors, the calorimeter system, and the MUON system.

The simulation of high-energy collisions, of the decays of the generated particles, and of the physics processes occurring within the detector by the decay products are a key necessity of analysis, typically for separating the signal from background sources or for selection efficiency studies.
The simulation software of the LHCb experiment is built upon two main projects named \gauss and \boole~\cite{Clemencic:2011zza}, both based on the \gaudi framework~\cite{Barrand:2001ny}.
The \gauss framework implements the so-called generation and simulation phases, while the \boole application is responsible for the digitization phase.
The first step of any simulation production is the \emph{generation} phase in which the high-energy collisions are simulated with Monte Carlo generators such as \pythia~\cite{Sjostrand:2007gs} and \evtgen~\cite{Lange:2001uf}. 
The output of the generation phase is the set of long-lived particles able to traverse partially or entirely, depending on the particle species, the LHCb spectrometer. 
The radiation-matter interactions occurring within the detector by the traversing long-lived particles are reproduced during the \emph{simulation} phase that aims to compute the energy deposited in the active volumes relying on \geant~\cite{Allison:2006ve}. 
Lastly, during the \emph{digitization} phase, the energy deposits are converted into raw data mimicking the data format used in the LHCb Data Acquisition pipeline.

During the LHC \runII, the simulation of physics events at LHCb has taken more than 80\% of the distributed computing resources available to the experiment, namely the pledged CPU time.
The experiment has just resumed data taking after a major upgrade and will operate with higher luminosity and trigger rates collecting data samples at least one order of magnitude larger than in the previous LHC runs.
Meeting the foreseen needs in Run 3 conditions using only the traditional strategy for simulation, namely \emph{detailed simulation}, will far exceed the pledged resources.
Hence, the LHCb Collaboration is making great efforts to modernize the simulation software stack~\cite{Mazurek:2021abc, Mazurek:2022tlu} and develop novel and faster simulation options~\cite{Rama:2019smm, Maevskiy:2019vwj, Ratnikov:2021uwg, Rogachev:2022hjg, Anderlini:2022ofl}.

\section{The fast and ultra-fast simulation paradigms}
\label{sec:ultra-fast-sim}
The \emph{detailed simulation} of the dynamics of the hadron collisions and the interaction of all primary and secondary particles with the detector materials is extremely expensive in terms of CPU time.
It is therefore no surprise that the computation of energy deposits performed by \geant consumes more than 90\% of the CPU resources spent by LHCb for simulation.

\begin{figure}[h!]
    \begin{minipage}{0.55\textwidth}
        \centering
        \includegraphics[width=\textwidth]{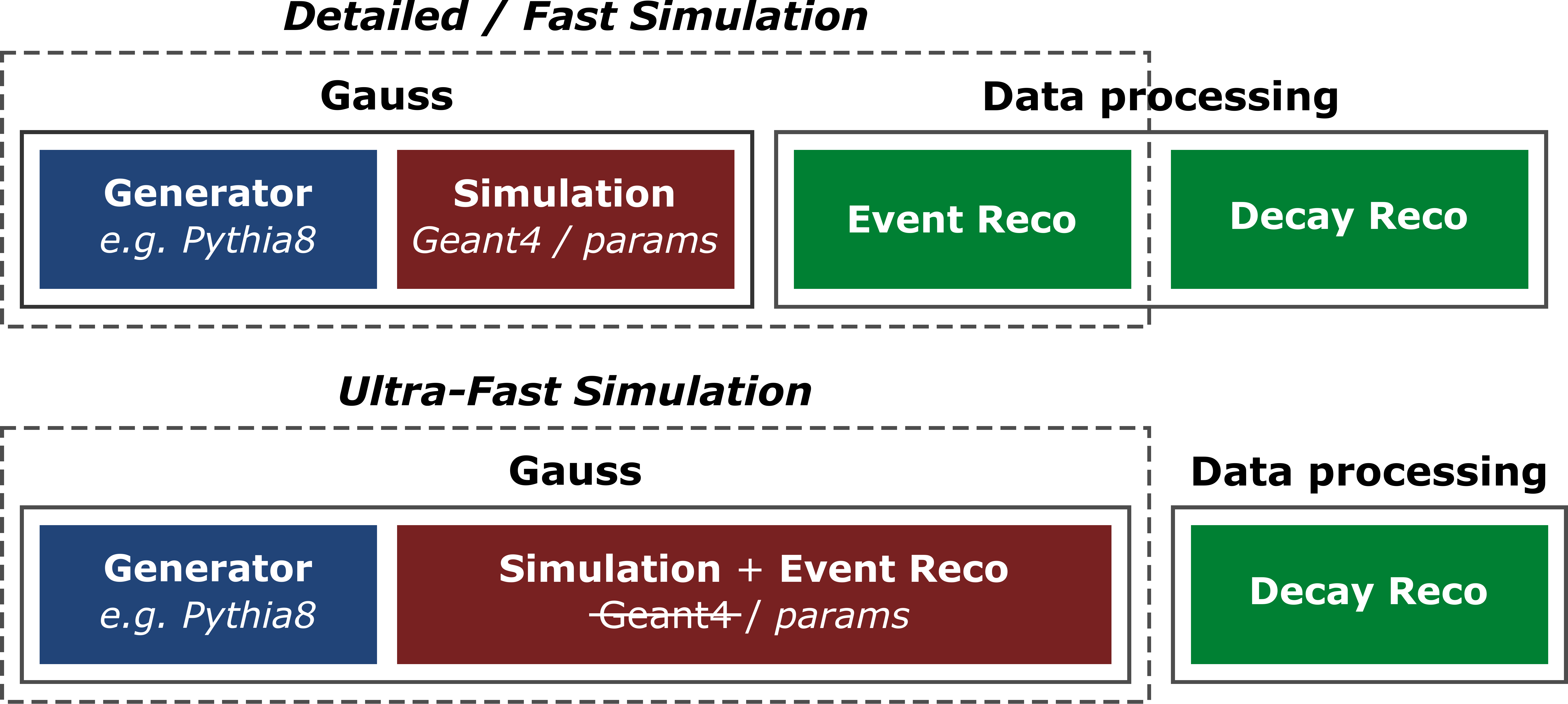}
    \end{minipage}
    \hskip -0.1\textwidth
    \begin{minipage}{0.55\textwidth}
        \caption{\label{fig:data-proc-flow}
            Schematic representation of the data processing flow in \emph{detailed} 
            and \emph{fast simulation} (top), and in \emph{ultra-fast simulation} (bottom).
        }
    \end{minipage}
\end{figure}

\newpage
Several strategies have been developed to reduce the computational cost of the simulation phase based on resampling techniques~\cite{Muller:2018vny} or parameterizations of energy deposits~\cite{Rama:2019smm, Ratnikov:2021uwg, Rogachev:2022hjg}.
These options offer cheaper alternative solutions to reproduce the low-level response of the LHCb detector and are typically named \emph{fast simulation} strategies.
The fast simulation options do not modify the traditional data processing flow described in Figure~\ref{fig:data-proc-flow}~(top), but rather allow to speed up the simulation phase up to a factor 20 with respect to the \emph{detailed simulation}.

A more radical approach is the one followed by the \emph{ultra-fast simulation} strategies which aim to parameterize directly the high-level response of the LHCb detector~\cite{Maevskiy:2019vwj, Anderlini:2022ofl}.
The core idea is to develop parameterizations able to transform generator-level particles information into reconstructed physics objects as schematically represented in Figure~\ref{fig:data-proc-flow}~(bottom). Such parameterizations can be built using \emph{deep generative models} that have proven to succeed in describing the response of the LHCb detector at different levels~\cite{Ratnikov:2023wof} and in offering reliable synthetic simulated samples~\cite{Anderlini:2022hgm, Anderlini:2022ckd}.

\section{Lamarr and its machine-learning-based parameterizations}
\label{sec:lamarr}
\lamarr~\cite{Anderlini:2022ofl} is a novel LHCb simulation framework implementing the \emph{ultra-fast simulation} paradigm. 
The \lamarr framework consists of a pipeline of modular parameterizations designed to take as input the particles generated by the event generators and provide as output high-level quantities representing the particles successfully reconstructed by LHCb. 
\lamarr is integrated with \gauss and disposes of a dedicated interface to the physics generators for selecting those particles that need to be propagated through the detector, splitting them into charged and neutral particles. 
The remainder of this document is devoted to discuss the implementation (this Section) and validation (Section~\ref{sec:validation}) of the pipeline currently provided by \lamarr for charged particles.

Most of the parameterizations used by \lamarr rely on machine learning algorithms that we can split into two main classes. 
The first class of models uses \emph{Gradient Boosted Decision Trees}~(GBDT) to parameterize efficiencies learning the fraction of candidates that are in acceptance, that have been successfully reconstructed or that have been selected as muons.
The second family of parameterizations is made up of \emph{Generative Adversarial Networks}~(GAN)~\cite{Goodfellow:2014} trained to reproduce the distributions of high-level physics quantities, typically conditioned~\cite{Mirza:2014} by the kinematics of the particles traversing a specific LHCb sub-detector.
Additional algorithms to define detector parameterizations are being explored, but currently are not part of the \lamarr pipeline~\cite{Graziani:2021vai, Mariani:2023rvl}.

Once taken the charged particles from physics generators, the first step performed by \lamarr is their propagation through the magnetic field following a trajectory approximated as two rectilinear segments with a single point of deflection (\emph{single $p_T$ kick} approximation). 
Then, the tracking acceptance and reconstruction efficiency are computed using GBDT models trained taking as input geometrical and kinematic features of the track.
The resulting tracks still have information at generator-level.
The promotion to high-level quantities, namely the application of the resolution effects due to, for example, multiple scattering phenomena, is carried out by GAN systems trained with \emph{binary cross-entropy} as loss function and equipped with \emph{skip connections}~\cite{He:2016}.
A similar GAN-based architecture is used to provide the correlation matrix obtained from the Kalman filter adopted in the reconstruction algorithm to define the position, slope and curvature of each track.

The LHCb PID system is parameterized using GAN-based models.
The high-level response of the RICH and MUON systems are reproduced using the particles kinematic information provided by the \lamarr tracking modules and a description of the detector occupancy, for example based on the total number of tracks traversing the detector.
The loss function adopted to train the PID-GAN models is the \emph{Wasserstein distance} where the Lipschitz constraint on the discriminator is enforced explicitly using a method called \emph{Adversarial Lipschitz Penalty}~(ALP) regularization~\cite{Terjek:2020}, resulting in WGAN-ALP models. 
GlobalPID classifiers, obtained in real data by combining RICH and MUON responses with information from the calorimeter system and features of the reconstructed tracks, are parameterized using similar GAN-based architectures that take as input what produced by the RICH-GAN and MUON-GAN models. 
Lastly, the efficiency of a binary muon-identification criterion, available since the earlier stage of data processing via a FPGA-based implementation, is parameterized with GBDT models.

Combining stacks of GBDT and GAN models, \lamarr provides the high-level response of the LHCb tracking and PID systems.
To validate the \emph{ultra-fast simulation} approach the chosen machine-learning-based models are trained on detailed simulated samples and the output of \lamarr is compared to the reference distributions as described in Section~\ref{sec:validation}.
An extension of the training procedure allows to train the PID models directly on real data (in particular on calibration samples~\cite{Aaij:2018vrk}), statistically subtracting any background components through weights application~\cite{Borisyak:2019vbz}.
The trained models are deployed through a transcompilation approach using the \texttt{scikinC} toolkit and dynamically linked to the \gauss application to ease the development and prototyping of new parameterizations~\cite{Anderlini:2022ltm}.

\section{Validation campaigns powered by \LbLcmunu decays}
\label{sec:validation}
As mentioned in the previous Section, the validation of the ultra-fast philosophy of \lamarr is based on the comparison between the distributions obtained from models trained on detailed simulation and the ones resulting from standard simulation strategies.
In particular, we discuss here the validation studies performed using simulated \LbLcmunu decays with \LcpKpi.
We are dealing with a semileptonic $\Lambda^0_b$ decay whose dynamics is not trivial and needs a faithful reproduction, highlighting the importance of interfacing to dedicated generators, in this case \evtgen.
This decay channel is being widely studied by LHCb, at the point that it is part of the calibration samples designed to provide data-driven corrections to the simulated PID efficiencies for proton candidates~\cite{Aaij:2018vrk}.
Interestingly, this $\Lambda^0_b$ decay includes in its final state the four charged particle species parameterized in the current version of \lamarr, namely muons, protons, kaons and pions.

\begin{figure}[b!]
    \centering
    \vskip -3mm
	\includegraphics[width=0.45\textwidth]{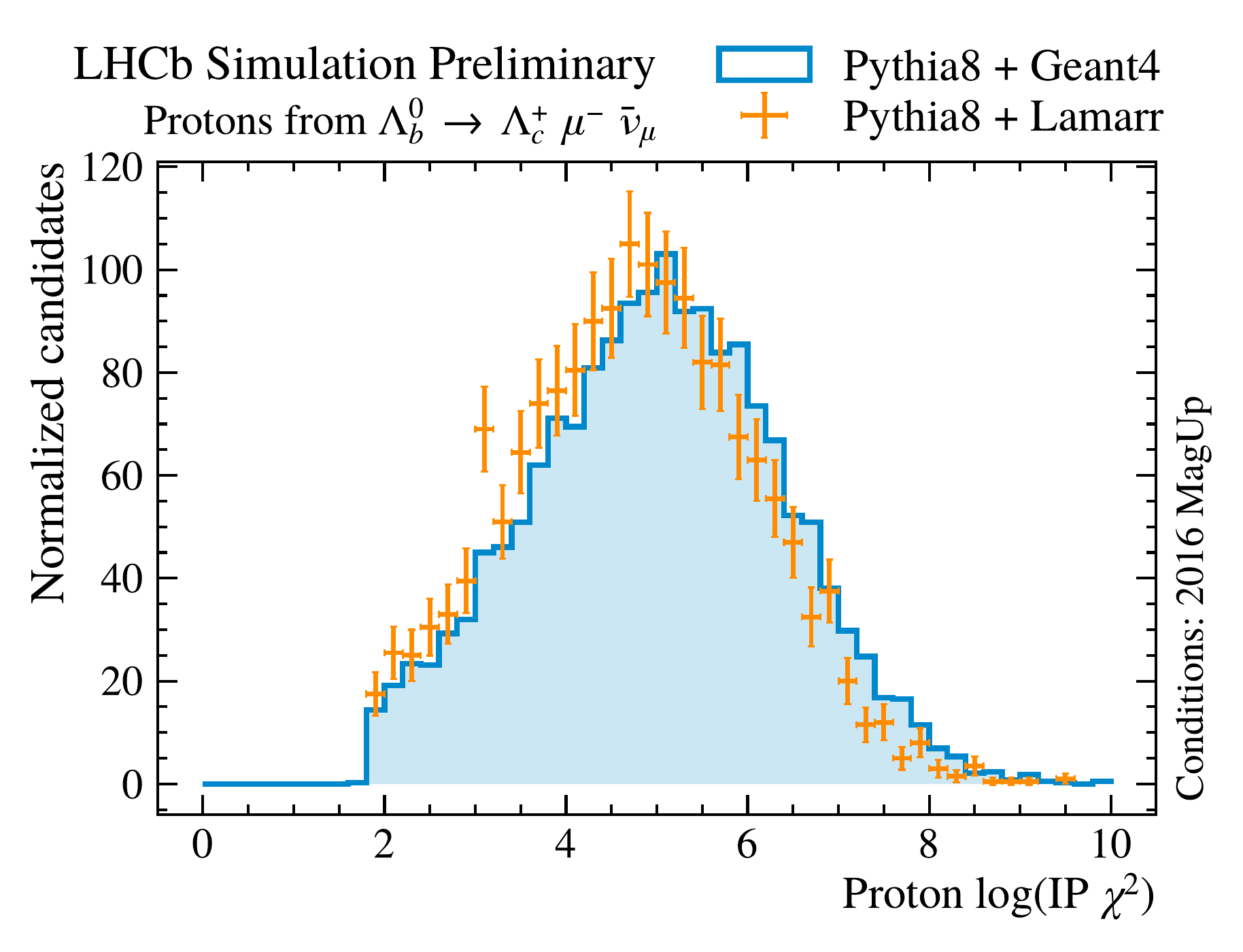}
    \includegraphics[width=0.45\textwidth]{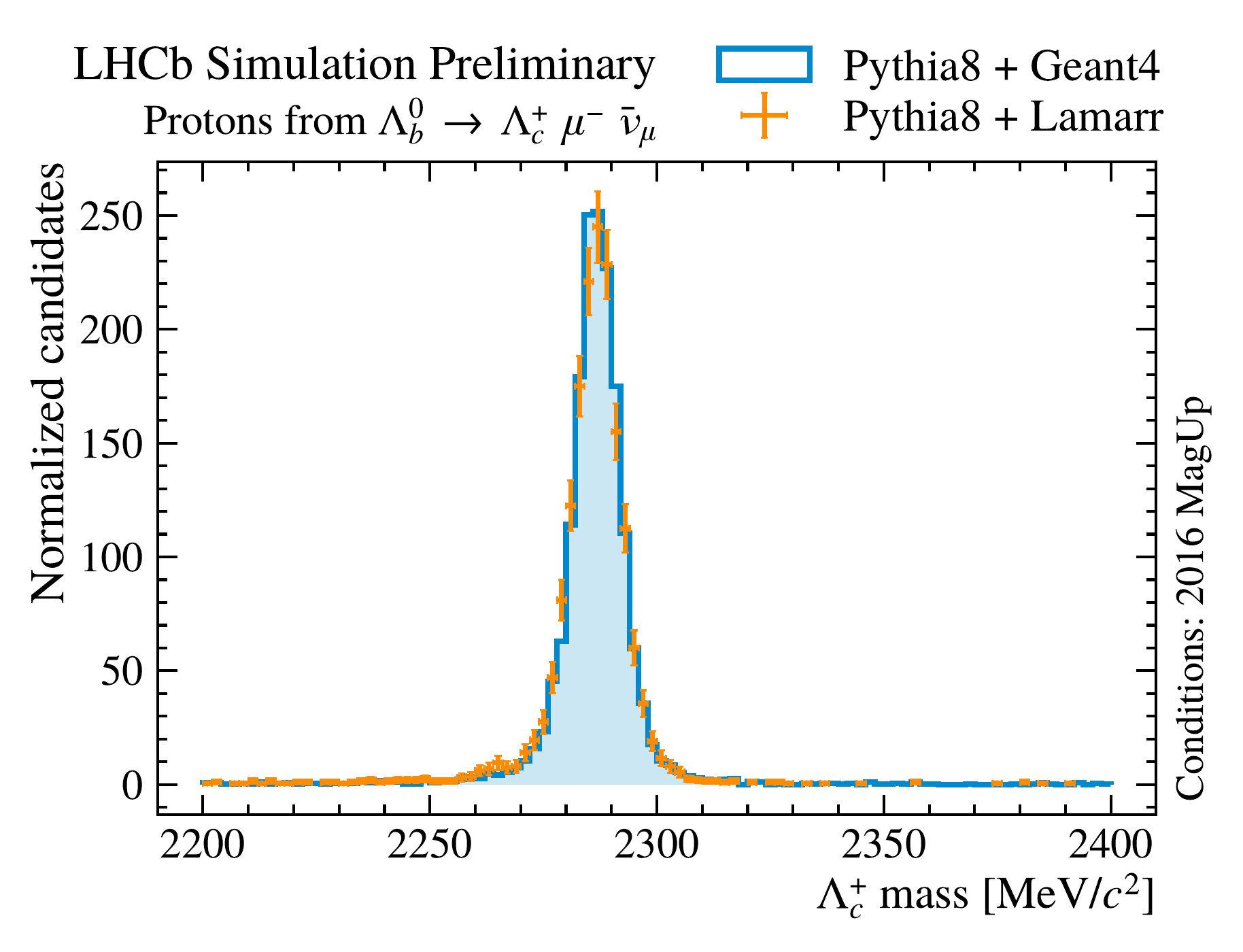}
    \includegraphics[width=0.45\textwidth]{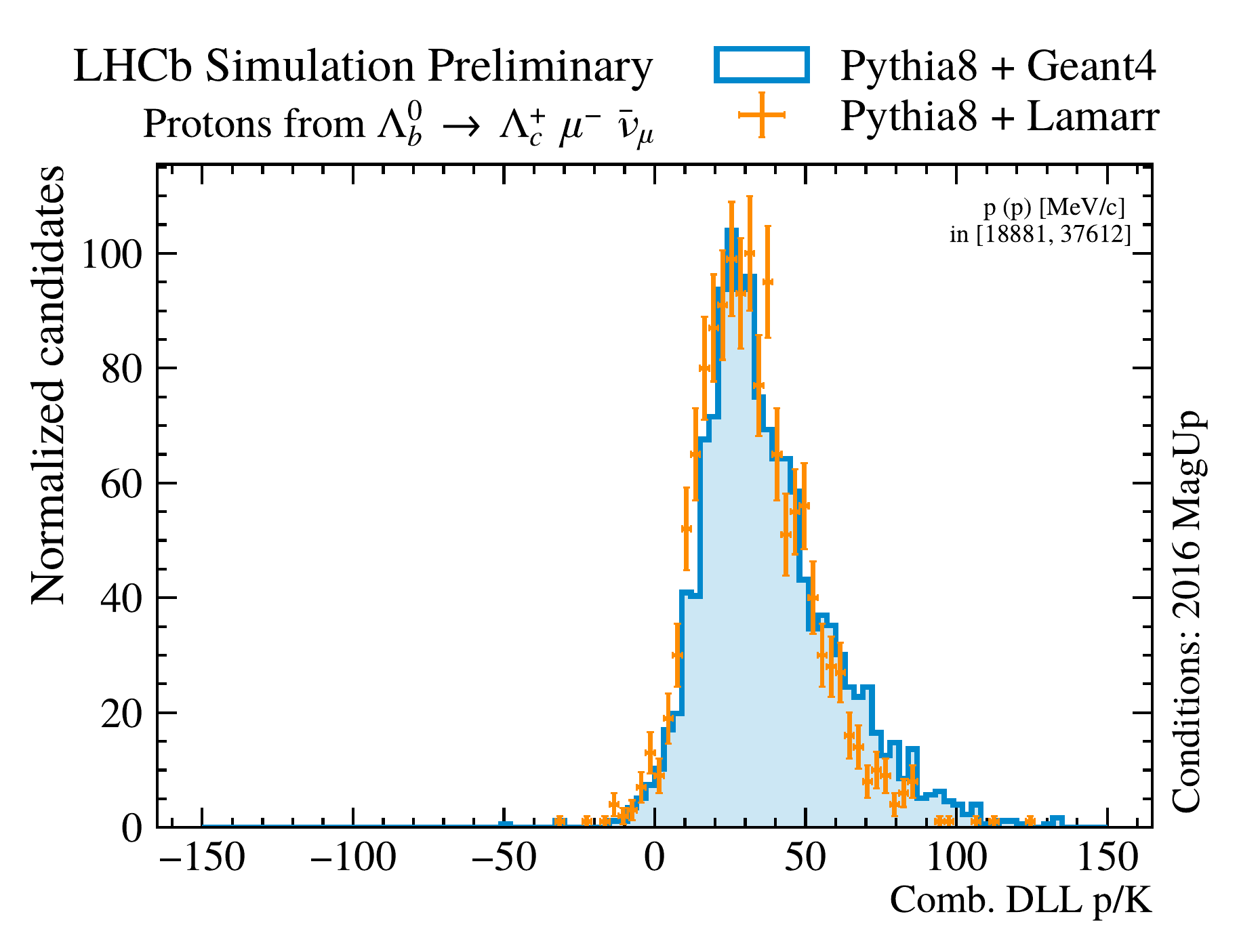}
	\includegraphics[width=0.45\textwidth]{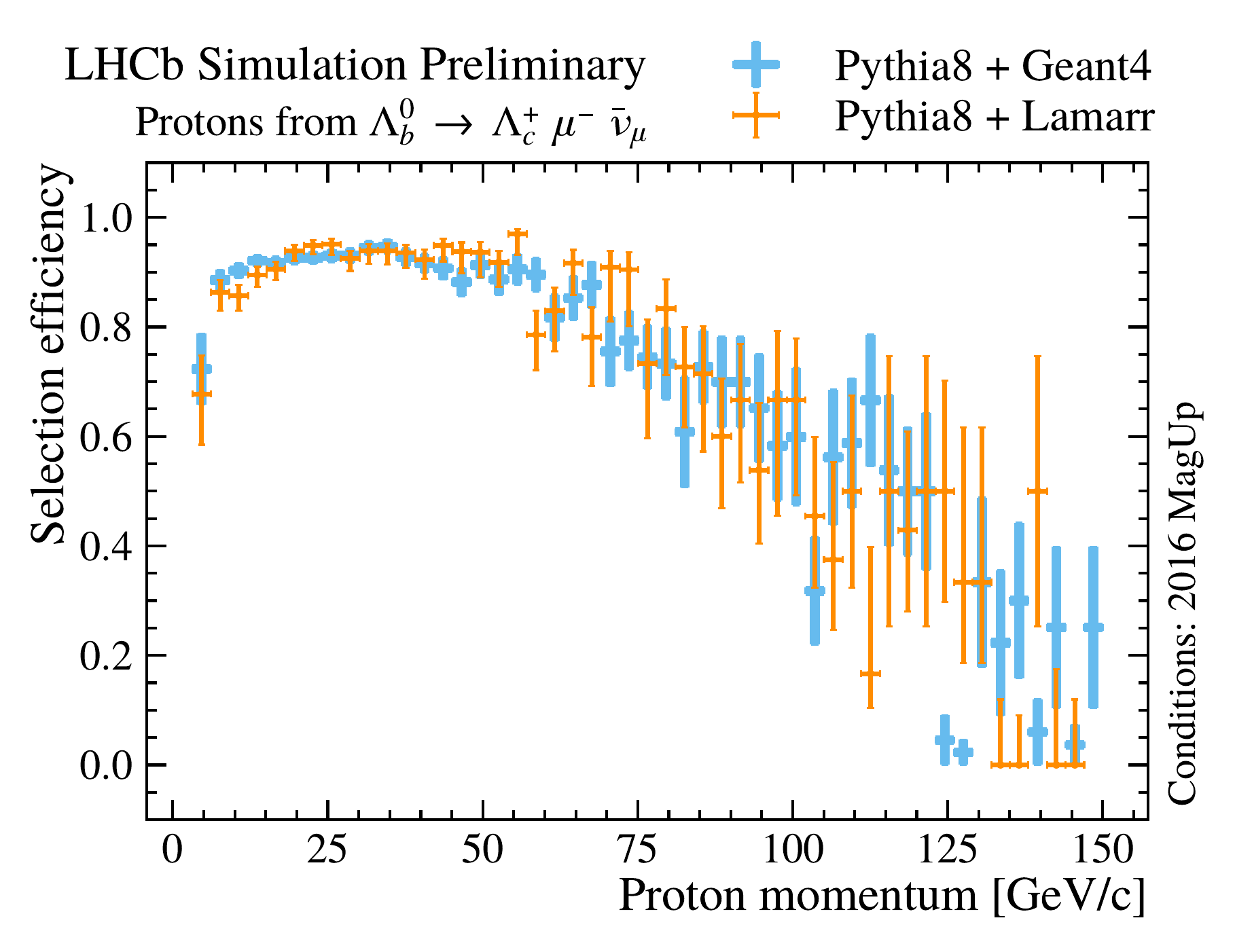}
    \vskip -2mm
	\caption{\label{fig:py8-res}
        Validation plots for \LbLcmunu decays with \LcpKpi simulated 
        with \pythia, \evtgen and \lamarr (orange markers) and compared 
        with \emph{detailed simulation} samples relying on \pythia, 
        \evtgen and \geant (cyan shaded histogram).
        Reproduced from \href{https://cds.cern.ch/record/2814081}{LHCB-FIGURE-2022-014}.
	}
    \vskip -3mm
\end{figure}

The validation of \lamarr tracking modules is reported in Figure~\ref{fig:py8-res}~(top) where a comparison between the distributions of the proton impact parameter $\chi^2$ (top left) and of the $\Lambda^+_c$ invariant mass (top right) are shown.
IP ${\chi^2}$ represents a measure of the inconsistency of the proton track with the PV obtained executing the same analysis algorithm both on \lamarr output and detailed simulated samples.
The agreement between the two invariant mass distributions proves that the decay dynamics is well reproduced and the resolution effects correctly parameterized.
To show the performance of the \lamarr PID parameterizations, the distribution of the Combined Differential Log-Likelihood (CombDLL) between the proton hypothesis and the kaon one on proton tracks is reported in Figure~\ref{fig:py8-res}~(bottom left) against what expected from detailed simulated samples.
A comparison between the selection efficiencies for a tight requirement on proton identification against pion hypothesis~(bottom right) is also shown in Figure~\ref{fig:py8-res}~(bottom right).

\section{Conclusion}
\label{sec:conclusion}
Developing new simulation techniques is an unavoidable requirement for LHCb to tackle the demand for simulated samples expected for Run 3 and those will follow.
The \emph{ultra-fast simulation} approach is a viable solution to reduce the pressure on pledged CPU resources and succeeds in describing the uncertainties introduced in the detection and reconstruction steps through the use of \emph{deep generative models}.
Such parameterization are provided to the LHCb software stack via the novel \lamarr framework, in which statistical models for tracking and charged particle identification have been deployed and validated with satisfactory results on \LbLcmunu decays.
Preliminary studies show that \lamarr is able to speed up the simulation production up to a factor 1000 with respect to \emph{detailed simulation}~\cite{Anderlini:2022ofl}.
Improvements on the quality of the parameterizations currently provided have been planned, relying on intense optimization campaigns on distributed computing resources~\cite{Barbetti:2023zce}.
Further development of the neutral particles pipeline is one of the major ongoing activities with the purpose of enhancing the variety of physics analyses that can benefits from \lamarr.

\section*{Acknowledgments}
This work is partially supported by ICSC -- \emph{Centro Nazionale di Ricerca in High Performance Computing, Big Data and Quantum Computing}, funded by European Union -- NextGenerationEU.

\section*{References}
\bibliographystyle{bib/iopart-num.bst}
\bibliography{bib/main.bib}

\end{document}